\documentclass[twocolumn]{IEEEtran}
\usepackage[utf8]{inputenc}
\usepackage[letterpaper,top=2cm,bottom=2cm,left=3cm,right=3cm,marginparwidth=1.75cm]{geometry}
\usepackage{hyperref}
\usepackage{graphicx}
\usepackage[compact]{titlesec}
\titlespacing{\section}{0pt}{*1}{*1}
\titlespacing{\subsection}{0pt}{*1}{*0}
\titlespacing{\subsubsection}{0pt}{*0}{*0}
\titleformat*{\paragraph}{\itshape}{}{}{}

\providecommand{\tightlist}{%
  \setlength{\itemsep}{0pt}\setlength{\parskip}{0pt}}
  
\title{Frugal Computing \\ {\LARGE On the need for low-carbon and sustainable computing and the path towards zero-carbon computing} }
\author{Wim Vanderbauwhede}
\date{November 2022}

\begin{document}

\maketitle
\begin{abstract}
 
 The current emissions from computing are almost 4\% of the world total. This is already more than emissions from the airline industry and are projected to rise steeply over the next two decades. By 2040 emissions from computing alone will account for more than half of the emissions budget to keep global warming below 1.5°C. Consequently, this growth in computing emissions is unsustainable.
 The emissions from production of computing devices exceed the emissions from operating them, so even if devices are more energy efficient producing more of them will make the emissions problem worse. Therefore we must extend the useful life of our computing devices.

As a society we need to start treating computational resources as finite and precious, to be utilised only when necessary, and as effectively as possible. 

We need \emph{frugal computing}: achieving our aims with less energy and material. 
 
\end{abstract}
\hypertarget{defining-computational-resources}{%
\section{Defining computational
resources}\label{defining-computational-resources}}

Computational resources are all resources of energy and material that
are involved in any given task that requires computing. For example,
when you perform a web search on your phone or participate in a video
conference on your laptop, the computational resources involved are
those for production and running of your phone or laptop, the mobile
network or WiFi you are connected to, the fixed network it connects to,
the data centres that perform the search or video delivery operations.
If you are a scientist running a simulator in a supercomputer, then the
computational resources involved are your desktop computer, the network
and the supercomputer. For an industrial process control system, it is
the production and operation of the Programmable Logic Controllers, IoT devices etc.

\hypertarget{computational-resources-are-finite}{%
\section{Computational resources are
finite}\label{computational-resources-are-finite}}

Since the start of general purpose computing in the 1970s, our society
has been using increasing amounts of computational resources.

For a long time the growth in computational capability as a function of
device power consumption has literally been exponential, a trend
expressed by
\href{https://www.britannica.com/technology/Moores-law}{Moore's law}.

With this growth in computational capability, increasing use of
computational resources has become pervasive in today's society. Until
recently, the total energy budget and carbon footprint resulting from
the use of computational resources has been small compared to the world
total. As a result, computational resources have until recently
effectively been treated as unlimited.

Because of this, the economics of hardware and software development have
been built on the assumption that with every generation, performance
would double for free. Now, this unlimited growth is no longer
sustainable because of a combination of technological limitations and
the climate emergency. Therefore, we need to do more with less.

Moore's law has effectively come to an end as integrated circuits can't
be scaled down any more. As a result, the improvement in performance per
Watt is slowing down continuously. On the other hand, the demand for
computational resources is set to increase considerably.

The consequence is that at least for the next decades, growth in demand
for computational resources will not be offset by increased power
efficiency. Therefore with business as usual, the total energy budget
and carbon footprint resulting from the use of computational resources
will grow dramatically to become a major contributor to the world total.

Furthermore, the resources required to create the compute devices and
infrastructure are also finite, and the total energy budget and carbon
footprint of production of compute devices is huge. Moore's Law has
conditioned us to doubling of performance ever two years, which has led
to very short effective lifetimes of compute hardware. This rate of
obsolescence of compute devices and software is entirely unsustainable.

Therefore, as a society we need to start treating computational
resources as finite and precious, to be utilised only when necessary,
and as frugally as possible. And as computing scientists, we need to
ensure that computing has the lowest possible energy consumption. And we
should achieve this with the currently available technologies because
the lifetimes of compute devices needs to be extended dramatically.

I would like to call this ``frugal computing'': achieving the same
results for less energy by being more frugal with our computing
resources.

\hypertarget{the-scale-of-the-problem}{%
\section{The scale of the problem}\label{the-scale-of-the-problem}}

\hypertarget{meeting-the-climate-targets}{%
\subsection{Meeting the climate
targets}\label{meeting-the-climate-targets}}

To limit global warming to 1.5°C, within the next decade a global
reduction from 55 gigatonnes C$_2$ equivalent (GtCO$_2$e) by 32 GtCO$_2$e to 23
GtCO$_2$e per year is needed \cite{key-7}. So by 2030 that would mean a
necessary reduction in overall CO$_2$ emissions of more than 50\%. By 2040,
a further reduction to 13 GtCO$_2$e per year is necessary. According to the
International Energy Agency \cite{key-12}, emissions from electricity are
currently estimated at about 10 GtCO$_2$e. 

The global proportion of
electricity from renewables is projected to rise from the current figure
of 22\% to slightly more than 30\% by 2040 \cite{key-17}. A more optimistic
scenario by the International Energy Agency \cite{key-19} projects 70\% of
electricity from renewables, but even in that scenario, generation from
fossil fuels reduces only slightly, so there is only a slight reduction
in emissions as a result.

In other words, we cannot count on renewables to eliminate CO$_2$ emissions
from electricity in time to meet the climate targets. The same is true for nuclear, offsetting and CCS: they will all come too late. Reducing the energy consumption is the only option.

\hypertarget{emissions-from-consumption-of-computational-resources}{%
\subsection{Emissions from consumption of computational
resources}\label{emissions-from-consumption-of-computational-resources}}

The consequence of the end of Moore's law was expressed most
dramatically in a 2015 report by the Semiconductor Industry Association
(SIA) ``Rebooting the IT Revolution: a call to action'' \cite{key-1},
which calculated that, based on projected growth rates and on the 2015
ITRS roadmap for CMOS chip engineering technologies \cite{key-18}

\begin{quote}
computing will not be sustainable by 2040, when the energy required for
computing will exceed the estimated world's energy production.
\end{quote}

It must be noted that this is purely the energy of the computing device,
as explained in the report. The energy required by e.g.~the data centre
infrastructure and the network is not included.

The SIA has reiterated this in their 2020 ``Decadal Plan for
Semiconductors'' \cite{key-2}, although they have revised the projection
based on a ``market dynamics argument'':

\begin{quote}
If the exponential growth in compute energy is left unchecked, market
dynamics will limit the growth of the computational capacity which would
cause a flattening out the energy curve.
\end{quote}

This is merely an acknowledgement of the reality that the world's energy
production is not set to rise dramatically, and therefore increased
demand will result in higher prices which will damp the demand. So
computation is not actually going to exceed the world's energy
production.

\begin{quote}
Ever-rising energy demand for computing vs.~global energy production is
creating new risk, and new computing paradigms offer opportunities to
dramatically improve energy efficiency.
\end{quote}

In the countries where most of the computational resources are consumed
(US and EU), electricity production accounts currently for 25\% of the
total emissions \cite{key-6}. According to the SIA's estimates, computation
accounts currently for a little less than 10\% of the total electricity
production but is set to rise to about 30\% by 2040. This would mean
that, with business as usual, computational resources would be
responsible for at least 10\% of all global CO$_2$ emissions by 2040.

The independent study ``Assessing ICT global emissions footprint: Trends
to 2040 \& recommendations'' \cite{key-3} corroborates the SIA figures: they
estimate the computing greenhouse gas emissions for 2020 between 3.0\%
and 3.5\% of the total, which is a bit higher than the SIA estimate of
2.5\% because it does take into account networks and datacentres. Their
projection for 2040 is 14\% rather than 10\%, which means a growth of 4x
rather than 3x.

To put it in absolute values, based on the above estimate, by 2040
energy consumption of compute devices would be responsible for 5 GtCO$_2$e,
whereas the target for world total emissions from all sources is 13
GtCO$_2$e.

 Other projections are of the same order, and the key message is that "computing’s carbon footprint growing at a rate unimaginable in other sectors." \cite{key-4}.
 
\hypertarget{emissions-from-production-of-computational-resources}{%
\subsection{Emissions from production of computational
resources}\label{emissions-from-production-of-computational-resources}}

To make matters worse, the carbon emissions resulting from the
production of computing devices exceeds those incurred during operation.
This is a crucial point, because it means that we can't rely on
next-generation hardware technologies to save energy: the production of
this next generation of devices will create more emissions than any
operational gains can offset. It does not mean research into more
efficient technologies should stop. But their deployment cycles should
be much slower. Extending the useful life of compute technologies must
become our priority.

The report about the cost of planned obsolescence by the European
Environmental Bureau \cite{key-9} makes the scale of the problem very clear.
For laptops and similar computers, manufacturing, distribution and
disposal account for 52\% of their
\href{https://www.sciencedirect.com/topics/earth-and-planetary-sciences/global-warming-potential}{Global
Warming Potential} (i.e.~the amount of CO$_2$-equivalent emissions caused).
For mobile phones, this is 72\%. The report calculates that the lifetime
of these devices should be at least 25 years to limit their Global
Warming Potential. Currently, for laptops it is about 5 years and for
mobile phones 3 years. According to \cite{key-10}, the
typical lifetime for servers in data centres is also 3-5 years, which
again falls short of these minimal requirements. According to this
paper, the impact of manufacturing of the servers is 20\% of the total,
which would require an extension of the useful life to 11-18 years.

\hypertarget{the-total-emissions-cost-from-computing}{%
\subsection{The total emissions cost from
computing}\label{the-total-emissions-cost-from-computing}}

\begin{figure}
    \centering
    \includegraphics[width=8cm]{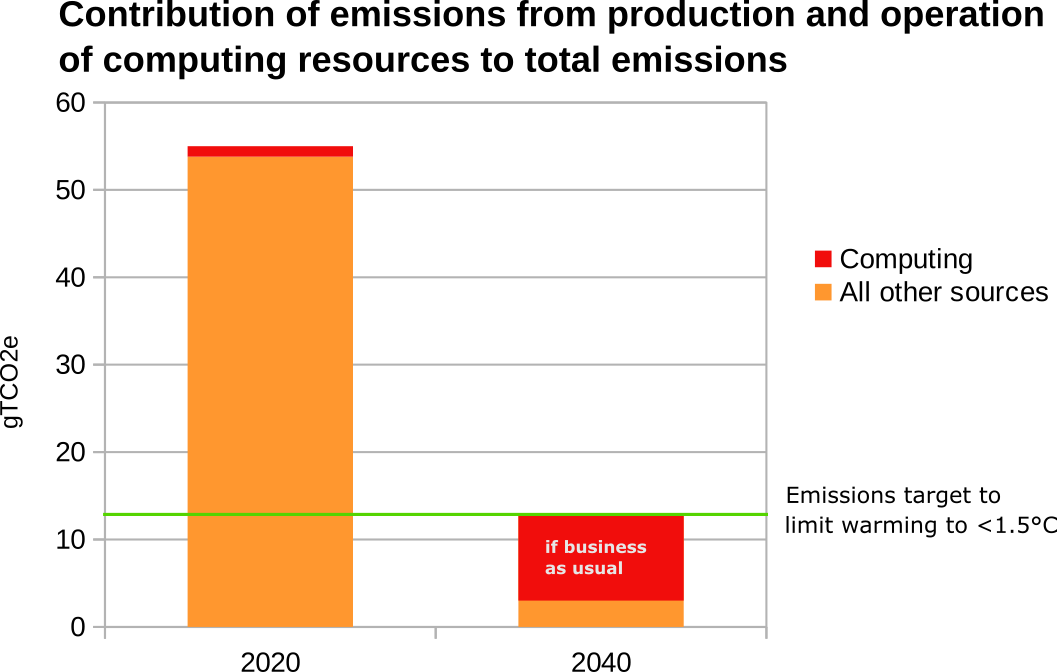}
    \caption{Actual and projected emissions from computing (production+operation), and 2040 emission target to limit warming to $\le$1.5ºC}
    \label{fig:actual-and-projected-emissions}
\end{figure}

Taking into account the carbon cost of both operation and production,
computing would be responsible for 10 GtCO$_2$e by 2040, almost 80\% of the
acceptable CO$_2$ emissions budget \cite{key-2,key-3,key-16}, as illustrated in Fig. \ref{fig:actual-and-projected-emissions}.

\hypertarget{a-breakdown-per-device-type}{%
\subsection{A breakdown per device
type}\label{a-breakdown-per-device-type}}

To decide on the required actions to reduce emissions, it is important
to look at the numbers of different types of devices and their energy
usage. If we consider mobile phones as one category, laptops and
desktops as another and servers as a third category, the questions are:
how many devices are there in each category, and what is their energy
consumption. The absolute numbers of devices in use are quite difficult
to estimate, but the yearly sales figures \cite{key-12} and estimates for the energy
consumption for each category \cite{key-13,key-14,key-15,key-16}
are readily available from various sources. The tables below show the
2020 sales (Table \ref{tab:number-of-device}) and yearly energy consumption estimates (Table \ref{tab:yearly-energy-consumption}) for each category of
devices. A detailed analysis is presented in
\cite{key-16}.

\begin{table}[!h]
    \centering
\begin{tabular}{ | c | c | }
\hline 
Device type & 2020 sales \\
\hline 
Phones &  3000M  \\ 
Servers &  13M  \\ 
Tablets &  160M  \\ 
Displays &  40M  \\ 
Laptops &  280M  \\ 
Desktops &  80M  \\ 
TVs & 220M  \\ 
IoT devices &  2000M \\
\hline 
\end{tabular}
    \caption{Number of devices sold worldwide in 2020}
    \label{tab:number-of-device}
\end{table}

The energy consumption of all communication and computation technology
currently in use in the world is currently around 3,000 TWh, about 11\%
of the world's electricity consumption, projected to rise by 3-4 times
by 2040 with business as usual according to
\cite{key-2}. This is a conservative estimate: the
study in \cite{key-16} includes a worst-case
projection of a rise to 30,000 TWh (exceeding the current world
electricity consumption) by 2030.

\begin{table}[]
    \centering
\begin{tabular}{ |c|c| }
\hline 
Device type & TWh \\
\hline 
TVs  &  560  \\
Other~Consumer Devices & 240  \\
Fixed~network (wired+wifi) &  1400  \\
Mobile~network &  100  \\
Data~centres &  700 \\
Total &  3000 \\
\hline 
\end{tabular}
    \caption{Yearly energy consumption estimates in TWh}
    \label{tab:yearly-energy-consumption}
\end{table}

The above data make it clear which actions are necessary: the main
carbon cost of phones, tablets and IoT devices is their production and
the use of the mobile network, so we must extend their useful life very
considerably and reduce network utilisation. Extending the life time is
also the key action for datacentres and desktop computers, but their
energy consumption also needs to be reduced considerably, as does the
energy consumption of the wired, WiFi and mobile networks.

\hypertarget{a-vision-for-low-carbon-and-sustainable-computing}{%
\section{A vision for low carbon and sustainable
computing}\label{a-vision-for-low-carbon-and-sustainable-computing}}

It is clear that urgent action is needed: in less than two decades, the
global use of computational resources needs to be transformed radically.
Otherwise, the world will fail to meet its climate targets, even with
significant reductions in other emission areas. The carbon cost of both
production and operation of the devices must be considerably reduced.

To use devices for longer, a change in business models as well as
consumer attitudes is needed. This requires raising awareness and
education but also providing incentives for behavioural change. And to
support devices for a long time, an infrastructure for repair and
maintenance is needed, with long-term availability of parts, open repair
manuals and training. To make all this happen, economic incentives and
policies will be needed (e.g.~taxation, regulation). Therefore we need
to convince key decision makers in society, politics and business.

\subsection*{A vision for zero-carbon computing}

Imagine that we can extend the useful life of our devices and even
increase their capabilities, purely through better computing science.
With every improvement, the computational capacity will in effect
increase without any increase in energy consumption. Meanwhile, we will
develop the technologies for the next generation of devices, designed
for energy efficiency as well as long life. Every subsequent cycle will
last longer, until finally the world will have computing resources that
last forever and hardly use any energy.

\begin{figure}
    \centering
    \includegraphics[width=8cm]{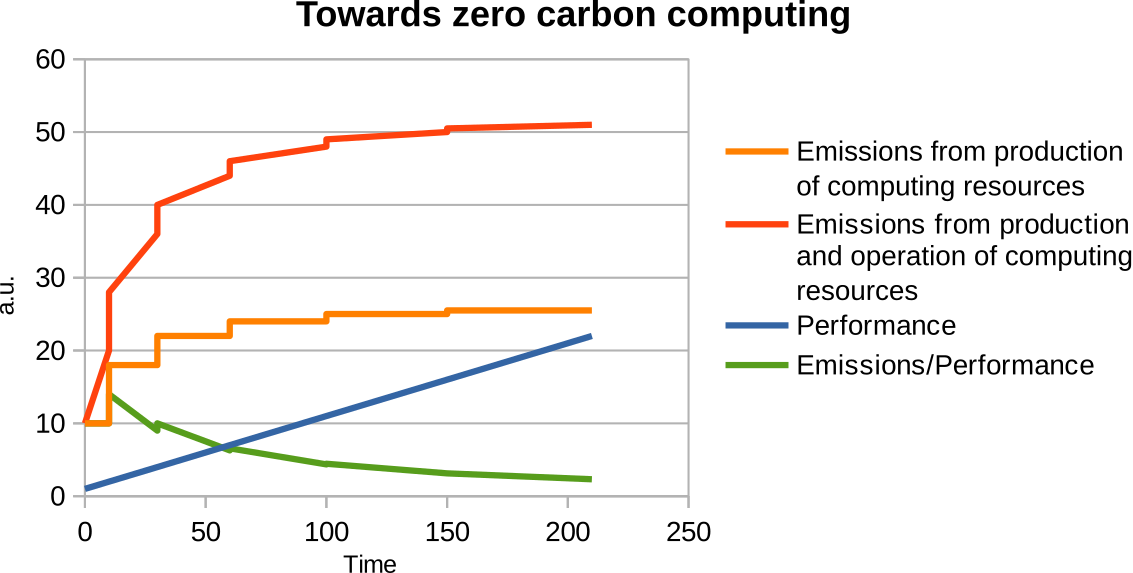}
    \caption{Towards zero carbon computing: increasing performance and lifetime and reducing emissions. Illustration with following assumptions: every new generation lasts twice as long as the previous one and cost half as much energy to produce; energy efficiency improves linearly with 5\% per year.}
    \label{fig:my_label}
\end{figure}

\hypertarget{research-challenges}{%
\section{Research challenges}\label{research-challenges}}

This is a very challenging vision, spanning all aspects of computing
science. To name just a few of the challenges:

\hypertarget{cloud-computing}{%
\subsection{Cloud computing}\label{cloud-computing}}

  Saving energy during operation, optimising for energy consumption
  (e.g.~DVFS, accelerators, scheduling and placement), more
  energy-efficient software on all layers;   better use of renewables through smarter scheduling.  Increasing the useful life, e.g reliability monitoring and
  early-warning systems, degradation-aware operation

\hypertarget{ultra-hd-video-vrar}{%
\subsection{Ultra-HD video \& VR/AR}\label{ultra-hd-video-vrar}}

Roll-out could lead to order of magnitude increases in video/3D traffic. To mitigate this, we need better compression (e.g.~tailored, AI), local rendering (e.g.~using FPGAs), better caching, energy-efficient edge computing

\hypertarget{iot}{%
\subsection{IoT}\label{iot}}

  The projected growth in IoT devices is huge, resulting in huge increase in
  network traffic as well as in emissions from production.  To mitigate this , we need to increase the device lifespan; reducing energy
  consumption helps primarily with this; and use edge computing to
  reduce network traffic.

\hypertarget{mobile-devices}{%
\subsection{Mobile devices}\label{mobile-devices}}

  The projected growth in mobile devices is still very large, and current
  lifespans much too short.  We mainly need longer-term software support, so better software engineering practices, in  particular relating to security.   Apps should be designed to minimise full-system energy consumption; user interfaces should nudge users towards energy efficient behaviour.

\hypertarget{research-directions}{%
\section{Research directions}\label{research-directions}}

We can also consider the research directions that would contribute to
this vision. Again, this is not a comprehensive list.

As a necessary condition, our systems must be as energy-efficient and long-lived as possible. This requires advances in many areas:
\begin{itemize}
\tightlist
\item
  Operating systems: energy-aware resource allocation and scheduling
\item
  Networking: Energy consumption as QoS criterion
\item
  Software engineering: better processes and sustainable practices will
  play a key role in extending lifetimes of systems.
\item
  Data centre/cloud: energy efficient heterogeneous systems, resource
  allocation, scheduling.
\end{itemize}

Sustainable systems need to be data-driven. Large systems produce huge
  amounts of system data. Making sense of this data is crucial for
  whole-system energy optimisation. 

 HCI has a key role to play in achieving low carbon computing
\begin{itemize}
\tightlist
 \item
  Make users aware of energy/carbon costs of their actions
\item
  Nudge user behaviour towards more sustainable practice
\item
  Human-computer interfaces influence both energy consumption and useful
  life of devices
\end{itemize}

Finally, formal methods also have a key role to play in many aspects of low-carbon computing:

\begin{itemize}
\tightlist
\item
  Programming languages need to become full-system energy aware.
\item
  Algorithms need to focus on minimising overall minimal energy consumption
\item
  Compilers need to compile for overall minimal energy consumption: not
  just CPU, also RAM, DMA, I/O wait etc; compilers need to be much better at selecting the right algorithms for given architectures, as algorithms strongly influence performance.
\end{itemize}

\section{Conclusion}

My position on the direction computing should take is clear: we should reduce emissions from computing by using less energy and less materials. From the computing science perspective, this provides us with the challenges of radical optimisation of energy efficiency and useful life of our computing resources. And this change needs to happen within the next two decades.

\end{document}